\documentclass[12pt,draftclsnofoot, onecolumn]{IEEEtran}

\usepackage{amssymb}
\usepackage[cmex10]{amsmath}
\usepackage{stfloats}
\usepackage{graphicx}
\usepackage{subfigure}
\usepackage{tabularx}
\usepackage{epsfig,epsf,color,balance,cite}
\usepackage{verbatim}
\usepackage{url}
\usepackage{bm}

\newtheorem{theorem}{\bf Theorem}
\newtheorem{proposition}{\bf Proposition}
\newtheorem{lemma}{\bf Lemma}

\usepackage{algorithm}
\usepackage{algorithmic}

\usepackage{color}
\definecolor{myc1}{rgb}{0,0,0}

\begin{document}

\title{\huge{Sum-Rate Maximization of Uplink Rate Splitting Multiple Access (RSMA) Communication}  }

\author{
\IEEEauthorblockN{Zhaohui Yang,
                  Mingzhe Chen,
                  Walid Saad, \IEEEmembership{Fellow, IEEE},
                  Wei Xu, \IEEEmembership{Senior Member, IEEE},
                  and Mohammad Shikh-Bahaei, \IEEEmembership{Senior Member, IEEE}
                  }
\thanks{A preliminary version of this work was submitted to IEEE Globecom 2019 \cite{rsma201902}.}
\thanks{Z. Yang  and M. Shikh-Bahaei are with the Centre for Telecommunications Research, Department of Informatics, King’s College London, WC2B 4BG, UK, Emails: yang.zhaohui@kcl.ac.uk, m.sbahaei@kcl.ac.uk.}
\thanks{M. Chen is with the Chinese University of Hong Kong, Shenzhen, China, and also with the Electrical Engineering Department of Princeton University, USA, Email: mzchen00@gmail.com.}
\thanks{W. Saad is with Wireless@VT, Bradley Department of Electrical and Computer Engineering, Virginia Tech, Blacksburg, VA, USA, Email: walids@vt.edu.}
\thanks{ W. Xu is with the National Mobile Communications Research
Laboratory, Southeast University, Nanjing 210096, China,  Email: wxu@seu.edu.cn.}
 }

\maketitle

\begin{abstract}
In this paper, the problem of maximizing the wireless users' sum-rate for uplink rate splitting multiple access (RSMA) communications is studied. In the considered model, each user transmits a superposition of two messages to a base station (BS) with separate transmit power and the BS uses a successive decoding technique to decode the received messages. To maximize each user's transmission rate, the users must adjust their transmit power and the BS must determine the decoding order of the messages transmitted from the users to the BS. This problem is formulated as a sum-rate maximization problem with proportional rate constraints by adjusting the users' transmit power and the BS's decoding order. However, since the decoding order variable in the optimization problem is discrete, the original maximization problem with transmit power and decoding order variables can be transformed into a problem with only the rate splitting variable. Then, the optimal rate splitting of each user is determined. Given the optimal rate splitting of each user and a decoding order, the optimal transmit power of each user is calculated. Next, the optimal decoding order is determined by an exhaustive search method. To further reduce the complexity of the optimization algorithm used for sum-rate maximization in RSMA, a user pairing based algorithm is introduced, which enables two users to use RSMA in each pair and also enables the users in different pairs to be allocated with orthogonal frequency. For comparisons, the optimal sum-rate maximizing solutions with proportional rate constraints are obtained in closed form for non-orthogonal multiple access (NOMA), frequency division multiple access (FDMA), and time division multiple access (TDMA). Simulation results show that RSMA can achieve up to 10.0\%, 22.2\%, and 83.7\% gains in terms of sum-rate compared to NOMA, FDMA, and TDMA.
\end{abstract}

\IEEEpeerreviewmaketitle

\section{Introduction}
Driven by the rapid development of advanced multimedia applications, next-generation wireless networks \cite{saad2019vision} must support high spectral efficiency and massive connectivity.
%
In consequence,
rate splitting multiple access (RSMA) has been recently proposed as an effective approach to provide more general and robust transmission framework compared to non-orthogonal multiple access (NOMA) \cite{liu2018non,vaezi2018multiple,7263349,Zhiguo2017Survey,Yang2018Power,8485386} and space-division multiple access (SDMA).
 However, implementing RSMA in wireless networks also faces
several challenges \cite{7470942} such as  decoding order design and resource management
for message transmission.

Recently, a number of existing works such as in \cite{7470942,4039650,7434643,7555358,mao2018rate,mao2019rate,8610154,ahmad2019interference,rahmati2019energy,7513415,7152864,yang2019optimization} have studied a number of problems related to the implementation of RSMA in wireless networks.  In \cite{7470942}, the authors outlined the opportunities and challenges of using RSMA for multiple input multiple output (MIMO) based wireless networks. The authors in \cite{4039650} developed a rate splitting algorithm for the maximization of users' data rates. The authors in \cite{7555358} developed an algorithm to optimize the users' sum-rate in downlink RSMA under imperfect channel state information (CSI). The authors in \cite{7555358} optimized users' sum-rate in downlink multi-user multiple input single output (MISO) systems under imperfect CSI. The work in \cite{mao2018rate} showed that RSMA can achieve better performance than NOMA and SDMA. In \cite{mao2019rate}, the application of linearly-precoded rate splitting is studied for multiple input single output (MISO) simultaneous wireless information and power transfer (SWIPT) broadcast channel systems.  The authors in \cite{8610154} investigated the rate splitting-based robust transceiver design problem in a multi-antenna interference channel with SWIPT under the norm-bounded errors of CSI. The work in \cite{ahmad2019interference} developed a transmission scheme that combines rate splitting, common message decoding, clustering and coordinated beamforming so as to maximize the weighted sum-rate of users. In \cite{rahmati2019energy}, the energy efficiency of the RSMA and NOMA schemes is studied in a downlink millimeter wave transmission scenario. The authors in \cite{7513415} used RSMA for a downlink multiuser MISO system with bounded errors of CIST. The data rate of using RSMA for two-receiver MISO broadcast channel with finite rate feedback is studied in \cite{7152864}. Our prior work in \cite{yang2019optimization} investigated the power management and rate splitting scheme to maximize the sum-rate of the users. However, most of the existing works such as in \cite{7470942,4039650,7434643,7555358,mao2018rate,mao2019rate,8610154,ahmad2019interference,rahmati2019energy,7513415,7152864,yang2019optimization} studied the use of RSMA for the downlink rather than in the uplink. In fact, using RSMA for uplink data transmission can theoretically achieve the optimal rate region \cite{485709}. Moreover, none of the existing works in \cite{7470942,4039650,7434643,7555358,mao2018rate,mao2019rate,8610154,ahmad2019interference,rahmati2019energy,7513415,7152864,yang2019optimization} jointly considered the optimization of power management and message decoding order for uplink RSMA. In practical RSMA deployments, the message decoding order will affect the transmission rate of the uplink users and, thus, it must be optimized.

The main contribution of this paper is a novel framework for optimizing power allocation and message decoding for uplink RSMA transmissions.  Our key contributions include:
\begin{itemize}
\item We consider the uplink of a wireless network that uses RSMA, in which each user transmits a superposition of two messages with different power levels and the base station (BS) uses a successive interference cancellation (SIC) technique to decode the received messages.
The power allocation and decoding order problem is formulated as an optimization problem whose goal is to maximize the sum-rate of all users under proportional rate constraints.
\item The non-convex sum-rate maximization problem with discrete decoding variable and transmit power variable is first transformed into an equivalent problem with only the rate splitting variable.
Then, the optimal solution of the rate splitting is obtained in closed form. Based on the optimal rate splitting of each user, the optimal transmit power can be derived under a given decoding order. Finally, the optimal decoding order is determined by exhaustive search.
To reduce the computational complexity, a low-complexity RSMA scheme based on user pairing is proposed to show near sum-rate performance of RSMA without user pairing.
\item We provide closed-form optimal solutions for sum-rate maximization problems in uplink NOMA, frequency division multiple access (FDMA), and time division multiple access (TDMA). Simulation results show that RSMA achieves better performance than NOMA, FDMA, and TDMA in terms of sum-rate.

\end{itemize}

The rest of this paper is organized as follows. The system model and problem formulation are described in Section \uppercase\expandafter{\romannumeral2}. The optimal solution is presented in Section \uppercase\expandafter{\romannumeral3}.
Section \uppercase\expandafter{\romannumeral4} presents a low-complexity sum-rate maximization scheme.
The optimal solutions of sum-rate maximization for NOMA, FDMA and TDMA are provided in Section \uppercase\expandafter{\romannumeral5}.
Simulation results are analyzed in Section \uppercase\expandafter{\romannumeral6}. Conclusions are drawn in Section \uppercase\expandafter{\romannumeral7}.




\section{System Model and Problem Formulation}
Consider a single cell uplink network with one BS serving a set $\mathcal K$ of $K$ users using RSMA.
In uplink RSMA,  each user first transmits a superposition code of two messages to the BS.
Then, the BS uses a SIC technique to decode the messages of all users \cite{485709}.

The transmitted message $s_k$ of user $k\!\in\!\mathcal K$ is given by:
\begin{equation}\label{sys1eq1}
s_k= \sum_{j=1}^2\sqrt{p_{kj}} s_{kj}, \quad \forall k \in\mathcal K,
\end{equation}
where $p_{kj}$ is the transmit power of message $s_{kj}$ from user $k$.

The total received message $s_0$ at the BS can be given by:
\begin{equation}\label{sys1eq2}
s_0=\sum_{k=1}^K \sqrt{h_k} s_k +n
=\sum_{k=1}^K \sum_{j=1}^2\sqrt{h_kp_{kj}}s_{kj}+n,
\end{equation}
where $h_k$ is the channel gain between user $k$ and the BS and $n$ is the additive white Gaussian noise.
Each user $k$ has a maximum transmission power limit $P_k$, i.e.,
$\sum_{j=1}^2 p_{kj} \leq P_k$.

To decode all messages $s_{kj}$ from the received message $s_0$, the BS will use SIC.
The decoding order at the BS is denoted by a permutation $\boldsymbol\pi$. The permutation $\boldsymbol\pi$ belongs to set $\Pi$ defined as the set
of all possible decoding orders of all $2K$ messages from $K$ users.
The decoding order of message $s_{kj}$ from user $k$ is ${\pi}_{kj}$.
The achievable rate of decoding message $s_{kj}$ is:
\begin{equation}\label{sys1eq3}
r_{kj} =
B \log_2 \left( 1+ \frac{h_kp_{kj}}{\sum_{\{(l\in\mathcal K,m\in\mathcal J)|{\pi}_{lm}>{ \pi}_{kj}\}} h_lp_{lm}+\sigma^2B}
\right),
\end{equation}
where $B$ is the bandwidth of the BS, $\sigma^2$ is the power spectral density of the Gaussian noise.
The set $\{(l\in\mathcal K,m\in\mathcal J)|{\pi}_{lm}>{\pi}_{kj}\}$ in \eqref{sys1eq3} represents the messages $s_{lm}$ that are decoded after message  $s_{kj}$.

Since the transmitted message of user $k$ includes messages $s_{k1}$ and $s_{k2}$, the achievable rate of user $k$ is given by:
\begin{equation}\label{sys1eq5}
r_k =\sum_{j=1}^2 r_{kj}.
\end{equation}

Our objective is to maximize the sum-rate of all users with  proportional rate constraints.
Mathematically, the sum-rate maximization problem can be formally posed as follows:
\begin{subequations}\label{ad2minr0}
\begin{align}
\mathop{\max}_{ \boldsymbol\pi,  \boldsymbol p} \quad& \sum_{k=1}^K r_k, \tag{\ref{ad2minr0}}  \\
\textrm{s.t.} \quad
&r_1:r_2:\cdots:r_K=D_1:D_2:\cdots D_K,\\
& \sum_{j=1}^2 p_{kj} \leq P_k, \quad \forall k \in \mathcal K,\\
&\boldsymbol\pi\in\Pi, p_{kj} \geq 0,\quad \forall k\in\mathcal K, j\in \mathcal J,
\end{align}
\end{subequations}
where $\boldsymbol p\!=\![p_{11},p_{12},\cdots,p_{K1},p_{K2}]^T$, $r_{k}$ is defined in \eqref{sys1eq5}, and $\mathcal J=\{1,2\}$.
$D_1, \cdots, D_K$ is a set of predetermined nonnegative values that are used to ensure proportional fairness among users.
The fairness index is defined as
\begin{equation}
\frac{\left(\sum_{k=1}^K D_k\right)^2}{K\sum_{k=1}^K D_k^2}
\end{equation}
 with the maximum value of 1 to be the greatest fairness case in which all users would achieve the same data rate \cite{shen2005adaptive}.
 With proper unitization, we set
\begin{equation}\label{ad2minr0eq0}
\sum_{k=1}^K D_k=1.
\end{equation}

Although it was stated in \cite{485709} that RSMA can reach the optimal rate region, no practical algorithm was proposed to compute the decoding order and power allocation. 
It is therefore necessary to quantify the uplink performance gains that RSMA can obtain compared to conventional multiple access schemes. 

Due to the non-linear equality constraint  (\ref{ad2minr0}a) and discrete variable $\boldsymbol \pi$,
 problem (\ref{ad2minr0}) is a non-convex mixed integer problem.
Hence, it is generally hard to solve problem (\ref{ad2minr0}).
Despite the non-convexity and discrete variable, we will next develop a novel algorithm to obtain the globally optimal solution to problem (\ref{ad2minr0}).

\section{Optimal Power Allocation and Decoding Order}
In this section, an effective algorithm is proposed to obtain the optimal power allocation and decoding order of sum-rate maximization problem \eqref{ad2minr0}.

\subsection{Optimal Sum-Rate Maximization}
Let $\tau$ be the sum-rate of all $K$ users. Given this new variable
$\tau$, problem \eqref{ad2minr0} can be rewritten as:
\begin{subequations}\label{ad2minr1}
\begin{align}
\mathop{\max}_{\tau,\boldsymbol\pi,  \boldsymbol p} \quad & \tau, \tag{\ref{ad2minr1}}  \\
\textrm{s.t.} \quad
&r_k = {D_k}\tau,\quad \forall k\in \mathcal K,\\
& \sum_{j=1}^2 p_{kj} \leq P_k, \quad \forall k \in \mathcal K,\\
&\boldsymbol\pi\in\Pi, p_{kj} \geq 0,\quad \forall k\in\mathcal K, j\in \mathcal J,
\end{align}
\end{subequations}
where $\tau$ is the sum-rate of all users since $\tau=\sum_{k=1}^K D_k \tau=\sum_{k=1}^K r_k$ according to (\ref{ad2minr0eq0}) and (\ref{ad2minr1}a).

Problem \eqref{ad2minr1} is challenging to solve due to the decoding order variable $\boldsymbol\pi$ with discrete value space.
To handle this difficulty, we provide the following lemma, which can be used for transforming problem \eqref{ad2minr1} into an equivalent problem without decoding order variable $\boldsymbol\pi$.

\begin{lemma}\label{ad2minr1_2le1}
In RSMA, under a proper decoding power order $\boldsymbol\pi$ and splitting power allocation $\boldsymbol p$, the optimal rate region can be fully achieved, i.e.,
\begin{equation}\label{ad2minr1eq1}
\sum_{k\in\mathcal K'}r_{k}   \leq B \log_2 \left( 1+ \frac{\sum_{k\in\mathcal K'}h_kP_{k}}{\sigma^2B}\right), \quad
 \forall \mathcal K'\subseteq \mathcal K \setminus \emptyset,
\end{equation}
where $\emptyset$ is an empty set and $\mathcal K'\subseteq \mathcal K\setminus \emptyset$ means that $\mathcal K'$ is a non-empty subset of $\mathcal K$.
\end{lemma}

Lemma \ref{ad2minr1_2le1} follows directly from \cite[Theorem 1]{485709}.
Based on Lemma \ref{ad2minr1_2le1}, we can use the rate variable to replace the power and decoding variables. In consequence,
problem \eqref{ad2minr1} can be equivalently transformed to
\begin{subequations}\label{ad2minr1_2}
\begin{align}
\mathop{\max}_{ \tau, \boldsymbol r} \quad & \tau, \tag{\ref{ad2minr1_2}}  \\
\textrm{s.t.} \quad
&  {r_{k} }={D_k}\tau ,\quad \forall k\in \mathcal K,\\
&  \sum_{k\in\mathcal K'}r_{k} \leq B \log_2 \left( 1+ \frac{\sum_{k\in\mathcal K'}h_kP_{k}}{\sigma^2B}\right), \quad \forall \mathcal K'\subseteq \mathcal K \setminus \emptyset,
\end{align}
\end{subequations}
where $\boldsymbol r= [r_{1}, r_{2}, \cdots,r_K]^T$.
In problem \eqref{ad2minr1_2}, the dimension of the variable is smaller than that in problem \eqref{ad2minr1}.
Moreover, the discrete decoding order variable is replaced by rate variable in problem \eqref{ad2minr1_2}.
Regarding the optimal solution of problem \eqref{ad2minr1_2}, we have the following lemma.

\begin{lemma}\label{ad2minr1_2le2}
For the optimal solution $(\tau^*, \boldsymbol r^*)$ of problem (\ref{ad2minr1_2}), there exists at least one $\mathcal K'\subseteq \mathcal K \setminus \emptyset$ such that $\sum_{k\in\mathcal K'}r_{k}^*
 = B \log_2 \left( 1+ \frac{\sum_{k\in\mathcal K'}h_kP_{k}}{\sigma^2B}\right)$.
\end{lemma}

\itshape  {Proof:}  \upshape
See Appendix A.
\hfill $\Box$

\begin{theorem}\label{ad2minr1_2th1}
The optimal solution of problem (\ref{ad2minr1_2}) is
\begin{equation}\label{ad2minr1_2eq2}
\tau^*=\min_{\mathcal  K'\subseteq \mathcal K \setminus \emptyset}\frac
{B \log_2 \left( 1+ \frac{\sum_{k\in\mathcal K'}h_kP_{k}}{\sigma^2B}\right)}{\sum_{k\in\mathcal K'}D_k},
\end{equation}
and
\begin{equation}\label{ad2minr1_2eq2_11}
r_k^*
= {D_k}{\min_{\mathcal  K'\subseteq \mathcal K \setminus \emptyset}\frac
{B \log_2 \left( 1+ \frac{\sum_{k\in\mathcal K'}h_kP_{k}}{\sigma^2B}\right)}{\sum_{k\in\mathcal K'}D_k}}, \quad \forall k \in \mathcal K.
\end{equation}
\end{theorem}

\itshape {Proof:}  \upshape
According to Lemma 2, there exists at least one $\mathcal K'\subseteq \mathcal K \setminus \emptyset$ such that
\begin{equation}
\sum_{k\in\mathcal K'}r_{k}^*
 = B \log_2 \left( 1+ \frac{\sum_{k\in\mathcal K'}h_kP_{k}}{\sigma^2B}\right).
\end{equation}
 To ensure the feasibility of  (\ref{ad2minr1_2}b), the optimal $\tau^*$ is given by \eqref{ad2minr1_2eq2}.
 Then, according to (\ref{ad2minr1_2}a), the optimal $r_k^*$ is determined as in (\ref{ad2minr1_2eq2_11}).
\hfill $\blacksquare$

From \eqref{ad2minr1_2eq2}, one can directly obtain the optimal sum-rate of problem \eqref{ad2minr1_2} in closed form, which can be helpful in characterizing the rate performance of RSMA.

Having obtained the optimal solution $(\tau^*,\boldsymbol r^*)$ of problem (\ref{ad2minr1_2}), we still need to calculate the optimal $(\boldsymbol\pi^*, \boldsymbol p^*) $ of the original problem (\ref{ad2minr1}).
Next, we introduce a new algorithm to obtain the optimal $(\boldsymbol\pi^*, \boldsymbol p^*) $ of  problem (\ref{ad2minr1}).

Substituting the optimal solution $(\tau^*,\boldsymbol r^*)$ of problem (\ref{ad2minr1_2}) into problem (\ref{ad2minr1}), we can obtain the following feasibility problem:
\begin{subequations}\label{ad2minr1_3}
\begin{align}
 \text{find} \quad & \boldsymbol\pi, \boldsymbol p,
 \tag{\ref{ad2minr1_3}} \\
\textrm{s.t.}\quad
& \sum_{j=1}^2\!B \log_2\! \left( \!\!1\!+\! \frac{h_kp_{kj}}{\sum_{\{(l\in\mathcal K,m\in\mathcal J)|{\pi}_{lm}\!>\!{ \pi}_{kj}\}} \!h_lp_{lm}\!+\!\sigma^2B}
\!\!\right)\!\!
=r_{k}^{*}, \quad\forall k\in \mathcal K,\\
& \sum_{j=1}^2 p_{kj} \leq P_k, \quad \forall k \in \mathcal K,\\
&\boldsymbol\pi\in\Pi, p_{kj} \geq 0,\quad \forall k\in\mathcal K, j\in \mathcal J.
\end{align}
\end{subequations}

Due to the decoding order constraint (\ref{ad2minr1_3}c), it is challenging to find the optimal solution of  problem \eqref{ad2minr1_3}.
To solve this problem, we first fix the decoding order $\boldsymbol\pi$ to obtain the power allocation and then exhaustively search $\boldsymbol\pi$.
Given decoding order $\boldsymbol\pi$, problem \eqref{ad2minr1_3} can be simplified as:
\begin{subequations}\label{ad2minr1_5}
\begin{align}
 \text{find} \quad & \boldsymbol p,
 \tag{\ref{ad2minr1_5}} \\
\textrm{s.t.}\quad
& \sum_{j=1}^2\!B \log_2\! \left( \!\!1\!+\! \frac{h_kp_{kj}}{\sum_{\{(l\in\mathcal K,m\in\mathcal J)|{\pi}_{lm}\!>\!{ \pi}_{kj}\}} \!h_lp_{lm}\!+\!\sigma^2B}
\!\!\right)\!\!
\geq r_{k}^{*}, \quad\forall k\in \mathcal K,\\
& \sum_{j=1}^2 p_{kj} \leq P_k, \quad \forall k \in \mathcal K,\\
& p_{kj} \geq 0,\quad \forall k\in\mathcal K, j\in \mathcal J.
\end{align}
\end{subequations}
Note that the equality in (\ref{ad2minr1_3}a) is replaced by the inequality in
(\ref{ad2minr1_5}a).
The reason is that any feasible solution to problem \eqref{ad2minr1_3} is also feasible to problem \eqref{ad2minr1_5}. Meanwhile, for a feasible solution to problem \eqref{ad2minr1_5}, we can always construct a feasible solution to problem \eqref{ad2minr1_3}.

To verify the feasibility of problem \eqref{ad2minr1_5}, we can construct the following problem by introducing a new variable $\alpha$:
\begin{subequations}\label{ad2minr1_6}
\begin{align}
\max_{\alpha,\boldsymbol p} \quad & \alpha,
 \tag{\ref{ad2minr1_6}} \\
\textrm{s.t.}\quad
& \sum_{j=1}^2\!B \log_2\! \left( \!\!1\!+\! \frac{h_kp_{kj}}{\sum_{\{(l\in\mathcal K,m\in\mathcal J)|{\pi}_{lm}\!>\!{ \pi}_{kj}\}} \!h_lp_{lm}\!+\!\sigma^2B}
\!\!\right)\!\!
\geq \alpha r_{k}^{*},\quad \forall k\in \mathcal K,\\
& \sum_{j=1}^2 p_{kj} \leq P_k, \quad \forall k \in \mathcal K,\\
& p_{kj}, \alpha \geq 0,\quad \forall k\in\mathcal K, j\in \mathcal J.
\end{align}
\end{subequations}

To show the equivalence between problems \eqref{ad2minr1_5} and \eqref{ad2minr1_6}, we provide the following lemma.
\begin{proposition}\label{ad2minr1_2le5}
Problem \eqref{ad2minr1_5} is feasible if and only if the optimal objective value $\alpha^*$ of problem \eqref{ad2minr1_6} is equal to or larger than 1.
\end{proposition}

\itshape {Proof:}  \upshape
On one side, if $\boldsymbol p$ is a feasible solution of problem \eqref{ad2minr1_5},  we can show that
$(\alpha=1,\boldsymbol p)$ is a feasible solution of problem \eqref{ad2minr1_6}, which indicates that the optimal objective value of problem \eqref{ad2minr1_6} should be equal to or larger than 1.

On the other side, if the optimal solution $(\alpha^*,\boldsymbol p^*)$ of problem \eqref{ad2minr1_6} satisfies $\alpha^*\geq 1$,
we can show that $\boldsymbol p^*$ is a feasible solution of problem \eqref{ad2minr1_5}.
\hfill $\Box$

Problem \eqref{ad2minr1_6} is non-convex due to constraints (\ref{ad2minr1_6}a).
To handle the non-convexity of (\ref{ad2minr1_6}), we adopt the difference of two convex function (DC) method, using which a non-convex problem  can be solved suboptimally by 
converting a non-convex problem into convex subproblems.
In order to obtain a near globally optimal solution of problem \eqref{ad2minr1_6}, we can try
multiple initial points $(\alpha, \boldsymbol p)$, which can lead to multiple locally optimal solutions.
Thus, a near globally optimal solution can be obtained
by choosing the locally optimal solution with the highest objective value among all locally optimal solutions.
To construct an initial feasible point, we first arbitrarily generate $\boldsymbol p$ that satisfies linear constraints (\ref{ad2minr1_6}b)-(\ref{ad2minr1_6}c), and then we set:
\begin{equation}
\alpha=\min_{k\in\mathcal K}\frac{\sum_{j=1}^2B \log_2 \left( 1+ \frac{h_kp_{kj}}{\sum_{\{(l\in\mathcal K,m\in\mathcal J)|{\pi}_{lm}>{ \pi}_{kj}\}} h_lp_{lm}+\sigma^2B}
\right) }{r_k^*}.
\end{equation}
By using the DC method,
the left hand side of (\ref{ad2minr1_6}a) satisfies:
\begin{eqnarray}\label{ad2minr1_6eq1}
\begin{aligned}
\:\:&\sum_{j=1}^2B \log_2 \left( 1+ \frac{h_kp_{kj}}{\sum_{\{(l\in\mathcal K,m\in\mathcal J)|{\pi}_{lm}>{ \pi}_{kj}\}} h_lp_{lm}+\sigma^2B}
\right)\nonumber\\
=
&\sum_{j=1}^2B \log_2 \!\left( \! {\sum_{\{(l\in\mathcal K,m\in\mathcal J)|{\pi}_{lm}\geq{ \pi}_{kj}\}} h_lp_{lm}\!+\!\sigma^2B}
\right)
-\!\sum_{j=1}^2B \log_2 \!\left( \!{\sum_{\{(l\in\mathcal K,m\in\mathcal J)|{\pi}_{lm}>{ \pi}_{kj}\}} h_lp_{lm}\!+\!\sigma^2B}
\!\right)
\end{aligned}
\end{eqnarray}
\begin{eqnarray}
\begin{aligned}
\nonumber
\geq&\sum_{j=1}^2B \log_2 \!\left( \! {\sum_{\{(l\in\mathcal K,m\in\mathcal J)|{\pi}_{lm}\geq{ \pi}_{kj}\}} h_lp_{lm}\!+\!\sigma^2B}
\right)
-\!\sum_{j=1}^2B \log_2\! \left(\! {\sum_{\{(l\in\mathcal K,m\in\mathcal J)|{\pi}_{lm}>{ \pi}_{kj}\}} h_lp_{lm}^{(n)}\!+\!\sigma^2B}\!
\right)\nonumber\\
&-\sum_{j=1}^2B\frac{\sum_{\{(l\in\mathcal K,m\in\mathcal J)|{\pi}_{lm}>{ \pi}_{kj}\}} h_l(p_{lm}-p_{lm}^{(n)})}
{(\ln 2)\sum_{\{(l\in\mathcal K,m\in\mathcal J)|{\pi}_{lm}>{ \pi}_{kj}\}} h_lp_{lm}^{(n)}+\sigma^2B}
\nonumber\\
\triangleq& r_{k,\text{lb}}(\boldsymbol p, \boldsymbol p^{(n)}),
\end{aligned}
\end{eqnarray}
where $p_{lm}^{(n)}$ represents the value of $p_{lm}$ at iteration $n$,
and the inequality follows from the fact that $\log_2(x)$ is a concave function and a concave function is always no greater than its first-order approximation.
By substituting the left term of constraints (\ref{ad2minr1_6}a) with the concave function $r_{k,\text{lb}}(\boldsymbol p, \boldsymbol p^{(n)})$, problem  (\ref{ad2minr1_6}) becomes convex, and can be effectively solved by the interior point method~\cite{boyd2004convex}.

\begin{algorithm}[t]
\caption{Optimal Sum-Rate Maximization for RSMA}
\begin{algorithmic}[1]
\STATE Obtain the optimal solution $(\tau^*,\boldsymbol r^*)$ of problem (\ref{ad2minr1_2}) according to Theorem 1.
\FOR{$\boldsymbol\pi\in\Pi$}
\FOR{$1:1:N$}
\STATE Arbitrarily generate a feasible solution $(\alpha^{(0)},\boldsymbol p^{(0)})$  of problem (\ref{ad2minr1_6}), and set $n=0$.
\REPEAT
\STATE
Obtain the optimal solution $(\alpha^{(n+1)},\boldsymbol p^{(n+1)})$ of  convex problem (\ref{ad2minr1_6}) by replacing the left term of constraints (\ref{ad2minr1_6}a) with $r_{k,\text{lb}}(\boldsymbol p, \boldsymbol p^{(n)})$.
\STATE Set $n=n+1$.
\UNTIL the objective value (\ref{ad2minr1_6}a) converges.
\ENDFOR
\STATE Obtain the optimal solution ($\alpha^*, \boldsymbol p^*$) of problem \eqref{ad2minr1_6} with the highest objective value.
\STATE If $\alpha^*\geq 1$,  break and jump to step 13.
\ENDFOR
\STATE Obtain the optimal decoding order $\boldsymbol\pi^*=\boldsymbol\pi$ and power allocation $\boldsymbol p^*$ of problem (\ref{ad2minr1_3}).
\end{algorithmic}
\end{algorithm}

The optimal sum-rate maximization algorithm for RSMA is provided in Algorithm 1, where $N$ is the number of initial points to obtain a near globally optimal solution of non-convex problem~(\ref{ad2minr1_6}).

\subsection{Complexity Analysis}

In Algorithm 1, the major complexity lies in solving problem (\ref{ad2minr1_2}) and problem \eqref{ad2minr1_3}.
To solve (\ref{ad2minr1_2}), from Theorem 1, the complexity is  $\mathcal O(2^K-1)$ since the set $\mathcal K$ has $2^K-1$ non-empty subsets.
According to steps 2-12, a near globally optimal solution of problem (\ref{ad2minr1_3}) is obtained via solving a series of convex  problems  with different initial points and decoding order strategies.
Considering that the dimension of variables in problem (\ref{ad2minr1_6}) is $1+2K$,
the complexity of solving convex problem  in  step 6  by using the standard
interior point method is $\mathcal O(K^3)$ \cite[Pages 487, 569]{boyd2004convex}.
Since the network consists of $K$ users and
each user transmits a superposition two messages (there are $2K$ messages in total), the decoding order set $\Pi$ consists of $(2K)!/2^K$ elements.
Given $N$ initial points, the total complexity of solving problem \eqref{ad2minr1_3} is $\mathcal O(NK^3(2K)!/2^K)$.
As a result, the total complexity of Algorithm 1 is $\mathcal O(2^K+NK^3(2K)!/2^K)$.

In practice, we consider small $K$ to reduce the SIC complexity, the computational complexity of Algorithm 1 can be practical.
To deal with a large number of users, the users can be classified into different groups with small number of users in each group. The users in different groups occupy different frequency bands and users in the same group are allocated to the same frequency band using RSMA \cite{Fang2016EE,7273963}.
For the special case with $K=2$, we can show that the optimal optimal decoding order and power allocation of problem \eqref{ad2minr1_3} can be obtained in closed form.
\subsection{RSMA with Two Users}

Based on Lemma 1, the rate region of RSMA with two users can be expressed by:
\begin{equation}
\{(r_1,r_2)|0\leq r_1 \leq R_1, 0\leq r_2\leq R_2, r_1+r_2\leq R_{\max}\},
\end{equation}
where
\begin{equation}\label{rsma2usersRate}
R_1=B\log_2\left(1+\frac{h_1P_1}{\sigma^2B}\right),
R_2=B\log_2\left(1+\frac{h_2P_2}{\sigma^2B}\right),
R_{\max}=B\log_2\left(1+\frac{h_1P_1+h_2P_2}{\sigma^2B}\right).
\end{equation}

According to Algorithm 1, the computational complexity needed to obtain the boundary point (as shown in Lemma 2 the optimal point to minimize always lies in the boundary point) of the rate region for RSMA is high.
In the following, we introduce a low-complexity method to obtain all boundary points of the rate region in two-user RSMA.

In two-user RSMA, only one user needs to transmit a superposition code of two messages and the other user transmits one message.
Without loss of generality, user 1 only transmits one message $s_{11}$, i.e., the transmit power for message $s_{12}$ is always 0.
\begin{lemma}\label{ad2sedle1}
For two-user RSMA, the optimal decoding order is $s_{21}$, $s_{11}$ and $s_{22}$.
For the boundary rate $(r_1,r_2)$, we consider the following three cases.

Case (1) $r_1=R_1$, $0\leq r_2 \leq R_{\max}-R_1$:
the optimal power allocation is
\begin{equation}
p_{11}=P_1, p_{12}=0, p_{21}= \frac{1}{h_2}\left( 2^{\frac{r_2}{B}}-1\right) (h_1P_1+\sigma^2B), p_{22}=0.
\end{equation}

Case (2) $r_2=R_2$, $0\leq r_1 \leq R_{\max}-R_2$:
the optimal power allocation is
\begin{equation}
p_{11}=\frac 1{h_1}{\left(2^{\frac{r_1}{B}}-1\right) (h_2P_2+\sigma^2B)}, p_{12}=0,
p_{21}=0,p_{22}=P_2.
\end{equation}


Case (3) $r_1+r_2=R_{\max}$, $0\leq r_1\leq R_1$, $0\leq r_2\leq R_2$:
the optimal power allocation is
\begin{equation}
p_{11}=P_1, p_{12}=0,
p_{21}=P_2-\frac{h_1P_1}{h_2\left(2^{\frac{r_1}{B}}-1\right)}+\frac{\sigma^2B}{h_2},
p_{22}=\frac{h_1P_1}{h_2\left(2^{\frac{r_1}{B}}-1\right)}-\frac{\sigma^2B}{h_2}.
\end{equation}
\end{lemma}

\itshape {Proof:}  \upshape
See  Appendix B.
\hfill $\Box$

\section{Low-Complexity Sum-Rate Maximization}
According to Section III-B, the computational complexity of sum-rate maximization for RSMA is extremely high.
In this section, we propose a low-complexity scheme for RSMA, where users are classified into different pairs\footnote{In this paper, we assume that the user pairing is given, which can be obtained according to matching theory \cite{Fang2016EE}  or  the order of channel gains \cite{7273963}.} and each pair consists of two users.
RSMA is used in each pair and different pairs are allocated with different frequency bands.
Assume that $K$ users are classified into $M$ pairs, i.e., $K=2M$.
The set of all pairs is denoted by $\mathcal M$.

For pair $m$, the allocated fraction of bandwidth is denoted by $f_m$.
Let $c_{mj}$ denote the data rate of user $j$ in pair $m$.
According to Lemma 1, we have:
\begin{equation}\label{low3minr1eq1}
c_{mj}\leq Bf_m \log_2 \left( 1+ \frac{ h_{mj}P_{mj}}{\sigma^2B f_m}\right), \quad \forall m \in \mathcal M, j \in \mathcal J,
\end{equation}
\begin{equation}\label{low3minr1eq2}
 c_{m1}+c_{m2}\leq B f_m \log_2 \left( 1+ \frac{ h_{m1}P_{m1}+h_{m2}P_{m2}}{\sigma^2B f_m}\right),\quad\forall m \in \mathcal M,
\end{equation}
where $h_{mj}$ denotes the channel gain between user $j$ in pair $m$ and the BS, and $P_{mj}$ is the maximal transmission power of user $j$ in pair $m$.

Similar to \eqref{ad2minr0}, the sum-rate maximization problem for RSMA with user pairing can be formulated as:
\begin{subequations}\label{low3minr1_0}
\begin{align}
\mathop{\max}_{   \boldsymbol f, \boldsymbol c } \quad & \sum_{m=1}^M \sum_{j=1}^2 c_{mj}, \tag{\theequation}  \\
\textrm{s.t.} \quad
&c_{11}:c_{12}:\cdots:c_{M2} =D_{11}:D_{12}:\cdots:D_{M2} \\
& \sum_{m=1}^M f_m=1,\\
&c_{mj}\leq Bf_m \log_2 \left( 1+ \frac{ h_{mj}P_{mj}}{\sigma^2B f_m}\right), \quad \forall m \in \mathcal M, j \in \mathcal J,\\
& c_{m1}+c_{m2}\leq B f_m \log_2 \left( 1+ \frac{ h_{m1}P_{m1}+h_{m2}P_{m2}}{\sigma^2B f_m}\right),\quad\forall m \in \mathcal M, \\
&f_{m},c_{m1},c_{m2}  \geq 0,\quad \forall m\in\mathcal M,
\end{align}
\end{subequations}
where $\boldsymbol f= [f_{1}, f_{2}, \cdots,f_{M}]^T$,
$\boldsymbol c= [c_{11}, c_{12}, \cdots, c_{M1},c_{M2}]^T$, and $D_{11}, D_{12}, \cdots, D_{M1}, D_{M2}$ is a set of predetermined nonnegative values that are used to ensure proportional fairness among users with $\sum_{m=1}^M \sum_{j=1}^2 D_{mj}=1$.

Introducing a new variable
$\tau$, problem \eqref{low3minr1_0} can be rewritten as:
\begin{subequations}\label{low3minr1}
\begin{align}
\mathop{\max}_{\tau,   \boldsymbol f, \boldsymbol c } \quad & \tau, \tag{\theequation}  \\
\textrm{s.t.} \quad
&c_{mj} = {D_{mj}}\tau ,\quad \forall m\in \mathcal M, j\in\mathcal J,\\
& \sum_{m=1}^M f_m=1,\\
&c_{mj}\leq Bf_m \log_2 \left( 1+ \frac{ h_{mj}P_{mj}}{\sigma^2B f_m}\right), \quad \forall m \in \mathcal M, j \in \mathcal J,\\
& c_{m1}+c_{m2}\leq B f_m \log_2 \left( 1+ \frac{ h_{m1}P_{m1}+h_{m2}P_{m2}}{\sigma^2B f_m}\right),\quad\forall m \in \mathcal M, \\
&f_{m},c_{m1},c_{m2} \geq 0,\quad \forall m\in\mathcal M.
\end{align}
\end{subequations}



To solve problem (\ref{low3minr1}), we can use the bisection method to obtain the optimal solution.
Denote the optimal objective value of problem (\ref{low3minr1}) by $\tau^*$.
\begin{figure}
\centering
\vspace{-1em}
\includegraphics[width=4in]{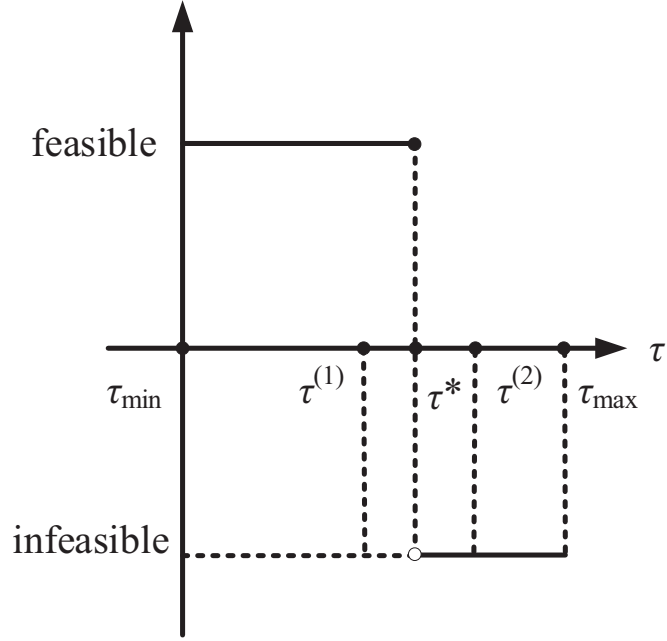}
\vspace{-3em}
\caption{An illustration of the bisection method.}\label{bisection2}
\vspace{-2em}
\end{figure}
We can conclude that problem (\ref{low3minr1}) is always feasible with $\tau<\tau^*$ and infeasible with $\tau>\tau^*$.
This motivates us to use the bisection method to find the optimal $\tau^*$, as shown in Fig. \ref{bisection2}, where $\tau^{(n)}$ is the value of $\tau$ in the $n$-th iteration and $[\tau_{\min},\tau_{\max}]$ is the initial value interval of $\tau$.
To show the feasibility of problem (\ref{low3minr1}) for each given $\tau$, we solve a feasibility problem with constraints (\ref{low3minr1}a)-(\ref{low3minr1}e).
With given $\tau$, the feasibility problem of (\ref{low3minr1}) becomes
\begin{subequations}\label{low3minr1_2}
\begin{align}
\text{find}\quad\: &{\boldsymbol f, \boldsymbol c } , \tag{\theequation}  \\
\textrm{s.t.} \quad
&c_{mj} = {D_{mj}}\tau ,\quad \forall m\in \mathcal M, j\in\mathcal J,\\
& \sum_{m=1}^M f_m=1,\\
&c_{mj}\leq Bf_m \log_2 \left( 1+ \frac{ h_{mj}P_{mj}}{\sigma^2B f_m}\right), \quad \forall m \in \mathcal M, j \in \mathcal J,\\
& c_{m1}+c_{m2}\leq B f_m \log_2 \left( 1+ \frac{ h_{m1}P_{m1}+h_{m2}P_{m2}}{\sigma^2B f_m}\right),\quad\forall m \in \mathcal M, \\
&f_{m},c_{m1},c_{m2} \geq 0,\quad \forall m\in\mathcal M.
\end{align}
\end{subequations}
Substituting  (\ref{low3minr1_2}a) into  (\ref{low3minr1_2}c) and (\ref{low3minr1_2}d),
we have:
\begin{equation}\label{low3minr1_2eq1}
 {D_{mj}}{\tau}\leq Bf_m \log_2 \left( 1+ \frac{ h_{mj}P_{mj}}{\sigma^2B f_m}\right),  \quad j\in\mathcal J,
\end{equation}
\begin{equation}\label{low3minr1_2eq2}
 ({D_{m1}\!+\!D_{m2}}){\tau}\!\leq\! B f_m \log_2 \!\left(\! 1\!+\! \frac{ h_{m1}P_{m1}\!+\! h_{m2}P_{m2}}{\sigma^2B f_m}\!\right).\!\!
\end{equation}
It can be proved that $g(x)=x\ln\left(1+\frac 1 x\right)$ is a monotonically increasing function.
Thus, to satisfy \eqref{low3minr1_2eq1} and \eqref{low3minr1_2eq2}, bandwidth fraction $f_m$ should satisfy:
\begin{equation}\label{low3minr1_2eq3}
f_m\geq \max\{f_{m1},f_{m2},f_{m3}\},
\end{equation}
where
\begin{equation}\label{low3minr1_2eq5}
f_{mk}=-\frac{(\ln2)D_{mk}h_{mk}P_{mk} }
{Bh_{mk}P_{mk}\tau W\left(-\frac{(\ln2)D_{mk}\sigma^2}{h_{mk}P_{mk}\tau} \text e^{-\frac{(\ln2)D_{mk}\sigma^2}{h_{mk}P_{mk}\tau}}
\right)+ {(\ln2)D_{mk}\sigma^2B}
}, \quad k=1,2,
\end{equation}
\begin{equation}\label{low3minr1_2eq6}
f_{m3}=-\frac{(\ln2)(D_{m1}+D_{m2})(h_{m1}P_{m1}+h_{m2}P_{m2}) }
{\beta+ {(\ln2)(D_{m1}+D_{m2})\sigma^2B}
},
\end{equation}
\begin{equation}
\beta=B(h_{m1}P_{m1}+h_{m2}P_{m2})\tau W\left(-\frac{(\ln2)(D_{m1}+D_{m2})\sigma^2}{(h_{m1}P_{m1}+h_{m2}P_{m2})\tau} \text e^{-\frac{(\ln2)(D_{m1}+D_{m2})\sigma^2}{(h_{m1}P_{m1}+h_{m2}P_{m2})\tau}}
\right),
\end{equation}
and $W(\cdot)$ is the Lambert-W function.


Based on \eqref{low3minr1_2eq3} and (\ref{low3minr1_2}b), we have:
\begin{equation}\label{low3minr1_2eq7}
\sum_{m=1}^M \max\{f_{m1},f_{m2},f_{m3}\} \leq 1.
\end{equation}

According to \eqref{low3minr1_2eq1}-\eqref{low3minr1_2eq7}, problem \eqref{low3minr1_2} has a feasible solution if and only if \eqref{low3minr1_2eq7} is satisfied.
As a result, the algorithm for obtaining the maximum sum-rate of problem \eqref{low3minr1_2} is summarized in Algorithm 2,
where $\tau^*$ is the optimal sum-rate of problem \eqref{ad2minr1_2}.

\begin{algorithm}[t]
\caption{: Low-Complexity Sum-Rate Maximization}
\begin{algorithmic}[1]
 \STATE Initialize $\tau_{\min}=0$, $\tau_{\max}=\tau^{*}$, and the tolerance $\epsilon$.
 \STATE Set $\tau=\frac{\tau_{\min}+\tau_{\max}}{2}$, and calculate $f_{m1}$, $f_{m2}$ and $f_{m3}$ according to (\ref{low3minr1_2eq5}) and (\ref{low3minr1_2eq6}), respectively.
\STATE Check the feasibility condition   (\ref{low3minr1_2eq7}). If problem (\ref{low3minr1_2}) is feasible, set $\tau_{\min}=\tau$. Otherwise, set $\tau=\tau_{\max}$.
\STATE If $(\tau_{\max}-\tau_{\min})/\tau_{\max}\leq \epsilon$, terminate. Otherwise, go to step 2.
\end{algorithmic}
\end{algorithm}

The complexity of the proposed Algorithm 2 in each step lies in checking the feasibility of problem (\ref{low3minr1_2}), which involves the complexity of $\mathcal O(M)$ according to (\ref{low3minr1_2eq5})-(\ref{low3minr1_2eq6}). 
As a result, the total complexity of the proposed Algorithm 2 is $\mathcal O( M\log_2(1/\epsilon))$, where $\mathcal O(\log_2(1/\epsilon))$ is the complexity of the bisection method with accuracy $\epsilon$.


\section{ Sum-Rate Maximization  for Uplink NOMA/FDMA/TDMA}
To evaluate the performance gain of the RSMA scheme proposed in Sections II-IV, and for comparison purposes, we will solve the  sum-rate maximization problems for  uplink NOMA, FDMA and TDMA schemes.

\subsection{NOMA}
Without loss of generality, the channel gains are sorted in descending order, i.e., $h_1\geq h_2\geq\cdots\geq h_K$.
In NOMA, the BS first decodes the messages of  users with high channel gains and then decodes the messages of users with low channel gains by subtracting the interference from decoded strong user.
The achievable rate of user $k$ with NOMA is calculated as \cite{saito2013non}:
\begin{equation}\label{sys1eq10}
r_{k}^{\text{NOMA}} =
B \log_2 \left( 1+ \frac{h_kq_k}{ \sum_{j=k+1}^Kh_jq_j+\sigma^2B}
\right),
\end{equation}
where $q_k$ is the transmit power of user $k$.
The transmission power $q_k$  has a maximum transmit power limit $P_k$, i.e., we have $q_k\leq P_k$, $\forall k \in \mathcal K$.

Similar to \eqref{ad2minr1}, the sum-rate maximization problem for uplink NOMA can be given by:
\begin{subequations}\label{ad2minn1}
\begin{align}
\mathop{\max}_{\tau,   \boldsymbol q} \quad & \tau, \tag{\ref{ad2minn1}}  \\
\textrm{s.t.} \quad
& B \log_2 \left( 1+ \frac{h_kq_k}{ \sum_{j=k+1}^Kh_jq_j+\sigma^2B}
\right)={D_k} \tau ,\quad \forall k\in \mathcal K,\\
&0\leq  q_{k} \leq P_k,\quad \forall k\in\mathcal K,
\end{align}
\end{subequations}
where $\boldsymbol q= [q_{1}, q_{2}, \cdots,q_{K}]^T$.
To obtain the optimal solution of problem \eqref{ad2minn1}, we provide the following theorem.
\begin{theorem}\label{ad2minn1th1}
The optimal solution of problem (\ref{ad2minn1}) is
 \begin{equation}\label{ad2minn1th1eq1}
 \tau^*=\min_{k\in\mathcal K} \tau_k,
 \end{equation}
 and
\begin{equation}\label{ad2minn1th1eq2} 
q_k^*=\frac 1 {h_k}{\left(2^{\frac{D_k\tau^*}{B}}-1\right)} \sum_{j=k+1}^K 2^{\frac{\sum_{l=k+1}^{j-1} D_l\tau^*}{B}}\left({2^{\frac{D_j\tau^*}{B}}} -1\right)\sigma^2B +\frac 1 {h_k}\left({2^{\frac{D_k\tau^*}{B}}} -1\right)\sigma^2B, \quad\forall k\in \mathcal K,
\end{equation}
where $\tau_k$ is the solution to
\begin{eqnarray} \label{ad2minn1th1eq3}
P_k=\frac 1 {h_k}{\left(2^{\frac{D_k\tau_k}{B}}-1\right)} \sum_{j=k+1}^K 2^{\frac{\sum_{l=k+1}^{j-1} D_l\tau_k}{B}}\left({2^{\frac{D_j\tau_k}{B}}} -1\right)\sigma^2B +\frac 1 {h_k}\left({2^{\frac{D_k\tau_k}{B}}} -1\right)\sigma^2B, \quad\forall k\in \mathcal K.
\end{eqnarray}
\end{theorem}

\itshape \textbf{Proof:}  \upshape
See  Appendix C.
\hfill $\Box$

Since the right hand side of \eqref{ad2minn1th1eq3} monotonically increases with $\tau_k$, the solution of $\tau_k$ to \eqref{ad2minn1th1eq3} can be effectively obtained by the bisection method.

\subsection{FDMA}
In FDMA, each user will be allocated a fraction of the BS bandwidth. Let $b_k$ denote the fraction of bandwidth allocated to user $k$. Then the data rate of user $k$ is:
\begin{equation}\label{sys1eq11}
r_{k}^{\text{FDMA}} =
Bb_k \log_2 \left( 1+ \frac{h_kP_k}{\sigma^2Bb_k}
\right).
\end{equation}
Note that user $k$ transmits with maximum power in \eqref{sys1eq11} since there is no inter-user interference and large power leads to high data rate.
Due to limited bandwidth, we have $\sum_{k=1}^K b_k =1$.

Similar to \eqref{ad2minr1}, the sum-rate maximization problem for uplink FDMA can be given by:
\begin{subequations}\label{ad2minf1}
\begin{align}
\mathop{\max}_{\tau,   \boldsymbol b} \quad & \tau, \tag{\ref{ad2minf1}}  \\
\textrm{s.t.} \quad
& Bb_k \log_2 \left( 1+ \frac{h_kP_k}{\sigma^2Bb_k}
\right)=D_k \tau,\quad \forall k\in \mathcal K,\\
& \sum_{k=1}^K b_k=1,\\
&b_{k} \geq 0,\quad \forall k\in\mathcal K,
\end{align}
\end{subequations}
where $\boldsymbol b= [b_{1}, b_{2}, \cdots,b_{K}]^T$.
Regarding the optimal solution of problem \eqref{ad2minf1}, we provide the following theorem.
\begin{theorem}\label{ad2minf1th1}
The optimal solution of problem (\ref{ad2minf1}) is $(\tau^*,\boldsymbol b^*)$, where $\tau^*$ is the solution of
 \begin{equation}\label{ad2minf1th1eq1}
-\sum_{k=1}^K \frac{(\ln2)D_kh_kP_k\tau }
{Bh_kP_k W\left(-\frac{(\ln2)D_k\tau\sigma^2}{h_kP_k} \text e^{-\frac{(\ln2)D_k\sigma^2\tau}{h_kP_k}}
\right)+ {(\ln2)D_k\sigma^2B\tau}
}=1
 \end{equation}
 and
\begin{equation} \label{ad2minf1th1eq2}
b_k^*=- \frac{(\ln2)D_kh_kP_k \tau^*}
{Bh_kP_k W\left(-\frac{(\ln2)D_k\sigma^2\tau^*}{h_kP_k} \text e^{-\frac{(\ln2)D_k\sigma^2\tau^*}{h_kP_k}}
\right)+ {(\ln2)D_k\sigma^2B\tau^*}
}.
 \end{equation}
\end{theorem}

\itshape \textbf{Proof:}  \upshape
See  Appendix D.
\hfill $\Box$

\subsection{TDMA}
In TDMA, each user will be assigned a fraction of time to use the whole BS bandwidth. Let $a_k$ be the fraction of time allocated to user $k$. The data rate of user $k$ is:
\begin{equation}\label{sys1eq12}
r_{k}^{\text{TDMA}} =
Ba_k \log_2 \left( 1+ \frac{h_kP_k}{\sigma^2B}
\right)
\end{equation}
with $\sum_{k=1}^K a_k =1$.

Similar to \eqref{ad2minr1}, the sum-rate maximization problem for uplink TDMA can be given by:
\begin{subequations}\label{ad2mint1}
\begin{align}
\mathop{\max}_{\tau,   \boldsymbol a} \quad & \tau, \tag{\ref{ad2mint1}}  \\
\textrm{s.t.} \quad
&  Ba_k \log_2 \left( 1+ \frac{h_kP_k}{\sigma^2B}
\right)=D_k \tau,\quad \forall k\in \mathcal K,\\
&\sum_{k=1}^K a_k=1,\\
&a_{k} \geq 0,\quad \forall k\in\mathcal K,
\end{align}
\end{subequations}
where $\boldsymbol a= [a_{1}, a_{2}, \cdots,a_{K}]^T$.
For problem \eqref{ad2mint1}, the optimal solution is given by the following theorem.
\begin{theorem}\label{ad2mint1th1}
The optimal solution of problem (\ref{ad2mint1}) is:
 \begin{equation}\label{ad2mint1th1eq1}
 \tau^*=\frac 1{\sum_{k=1}^K\frac{ D_k  }
{ B\log_2 \left( 1+ \frac{h_kP_k}{\sigma^2B}
\right)
}},
 \end{equation}
 and
\begin{equation}\label{ad2mint1th1eq2} 
a_k^*=\frac{ D_k  \tau^*}
{ B\log_2 \left( 1+ \frac{h_kP_k}{\sigma^2B}
\right)
}, \quad\forall k\in \mathcal K.
\end{equation}
\end{theorem}

\itshape \textbf{Proof:}  \upshape
See  Appendix E.
\hfill $\Box$

\subsection{Analysis and Discussion}
We define the optimal uplink sum-rates for RSMA, NOMA, FDMA and TDMA as $\tau^{\text{RSMA}}$, $\tau^{\text{NOMA}}$, $\tau^{\text{FDMA}}$ and $\tau^{\text{TDMA}}$, respectively.
For the optimal sum-rate with various uplink multiple access  schemes, we can state the following lemma.

\begin{lemma}
$\tau^{\text{RSMA}} \geq\tau^{\text{NOMA}}$
and
$\tau^{\text{RSMA}} \geq  \tau^{\text{FDMA}}\geq\tau^{\text{TDMA}}$.
\end{lemma}

Lemma 4 can be easily proved by the fact that for any feasible solution to NOMA/FDMA scheme, we can construct a feasible solution to RSMA with the same or better objective value and for any feasible solution to TDMA scheme,  we can construct a feasible solution to FDMA with the same or better objective value.
In order to illustrate the sum-rate performance of different multiple access schemes, we consider the special case with two users, i.e., $K=2$.

Denote $\mathcal R_2^{\text{X}}$ as the rate region of two users with multiple access $\text{X}\in\{\text{RSMA},\text{NOMA},\text{FDMA},$ $\text{TDMA}\}$.
Based on Lemma 2, the rate region of RSMA with two users can be expressed by:
\begin{equation}
\mathcal R_2^{\text{RSMA}}=\{(r_1,r_2)|0\leq r_1 \leq R_1, 0\leq r_2\leq R_2, r_1+r_2\leq R_{\max}\},
\end{equation}
where $R_1$, $R_2$ and $R_{\max}$ are defined in \eqref{rsma2usersRate}.

%
%


\begin{lemma}\label{ad2sedle2}
For two-user NOMA/FDMA/TDMA, we have
\begin{equation}\label{ad2sedle2eq1}
\mathcal R_2^{\text{NOMA}}=\left\{(r_1,r_2)|0\leq r_1\leq R_1, 0\leq r_2\leq R_2, r_2+B\log_2\left( { 2^{\frac{r_1}{B}}-1 }
\right)\leq B\log_2\left(
\frac{h_1 P_1}{\sigma^2B}
\right)\right\},
\end{equation}

\begin{equation}\label{ad2sedle2eq2}
\mathcal R_2^{\text{FDMA}}=\left\{(r_1,r_2)|0\leq r_1\leq R_1, 0\leq r_2\leq R_2,
f_1(r_1)+f_2(r_2)\leq 1,
  \right\},
\end{equation}
and
\begin{equation}\label{ad2sedle2eq3}
\mathcal R_2^{\text{TDMA}}=\left\{(r_1,r_2)|0\leq r_1\leq R_1, 0\leq r_2\leq R_2,
\frac{r_1}{R_1}+\frac{r_2}{R_2}\leq 1
  \right\},
\end{equation}
where
\begin{equation}\label{ad2sedle2eq5}
f_k(r_k)=-\frac{(\ln2) h_kP_k r_k }
{Bh_kP_k  W\left(-\frac{(\ln2)\sigma^2r_k}{h_kP_k } \text e^{-\frac{(\ln2)\sigma^2r_k}{h_kP_k }}
\right)+ {(\ln2)\sigma^2Br_k}
}, \quad k=1,2.
\end{equation}
\end{lemma}

\itshape \textbf{Proof:}  \upshape
See  Appendix F.
\hfill $\Box$

\begin{figure}[t]
\centering
\includegraphics[width=4.5in]{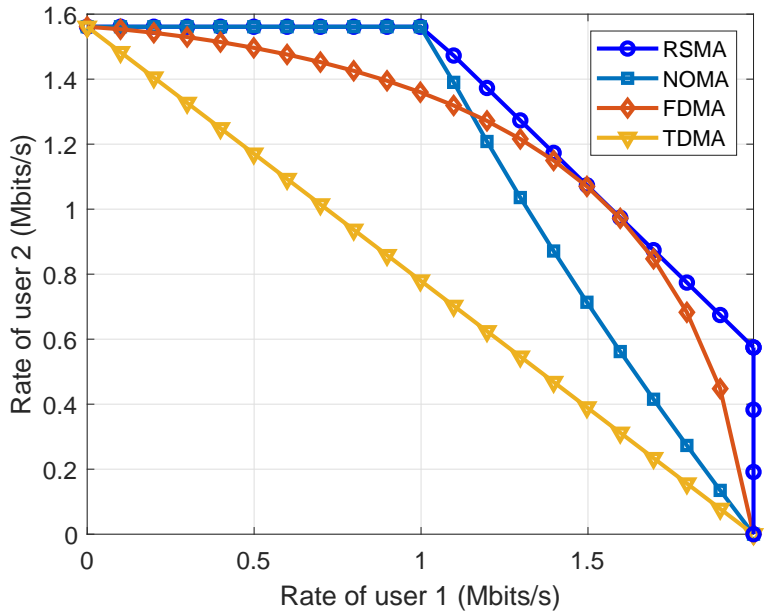}
\caption{An example of rate region for  RSMA, NOMA, FDMA and TDMA with $K=2$ users, $h_1=9.45\times 10^{-9}$, $h_2=6.17\times 10^{-9}$, $P_1=P_2=1$ dBm, $\sigma^2$= -174 dBm/Hz, and $B=1$ MHz.} \label{fig1}
\end{figure}

Based on Lemmas \ref{ad2sedle1} and \ref{ad2sedle2}, we provide an example of the rate region for multiple access schemes, as shown in Fig. \ref{fig1}.
From Fig.~\ref{fig1}, it is observed that RSMA has the largest rate region, while TDMA has the smallest rate region.
According to the proportional rate constraints (\ref{ad2minr0}a), we have $\frac{r_1}{r_2}= \frac{D_1}{D_2}$, which is a linear function through the original point in the rate coordinate $(r_1,r_2)$.
Thus, the optimal rate point can be obtained by finding the intersection between the rate region and the line  $r_2 =\frac{D_2}{D_1} r_1$.
Thus, it is shown from Fig. \ref{fig1} that RSMA has the best sum-rate performance and TDMA has the worst sum-rate performance, which verifies the theoretical findings in Lemma 5.

\section{Numerical Results}

\begin{table}[t]
\centering
\caption{System  Parameters} \label{tab:complexity}
\begin{tabular}{ccc}
  \hline
  \hline
  Parameter &   Value \\ \hline
Bandwidth of the BS $B$ &1 MHz  \\
Noise power spectral density $\sigma^2$& -174 dBm/Hz \\
Path loss model& $128.1+37.6\log_{10} d$ ($d$ is in km)\\
Standard deviation of shadow fading & $8$ dB\\
Maximum transmit power $P$ & 1 dBm\\
  \hline
  \hline
\end{tabular}
\end{table}

For our simulations, we deploy $K$ users uniformly in a square area of size $500$ m $\times$ $500$~m with the BS located at its center.
The path loss model is $128.1+37.6\log_{10} d$ ($d$ is in km)
and the standard deviation of shadow fading is $8$ dB.
In addition, and the noise power spectral density is  $\sigma^2=-174$ dBm/Hz.
Unless specified otherwise, we choose an equal maximum transmit power $P_1=\cdots=P_K=1$ dBm,  and a bandwidth $B=1$ MHz.
The main system parameters are summarized in Table I.
All statistical results are averaged over a large number of independent runs.

\begin{figure}[t]
\centering
\includegraphics[width=4.5in]{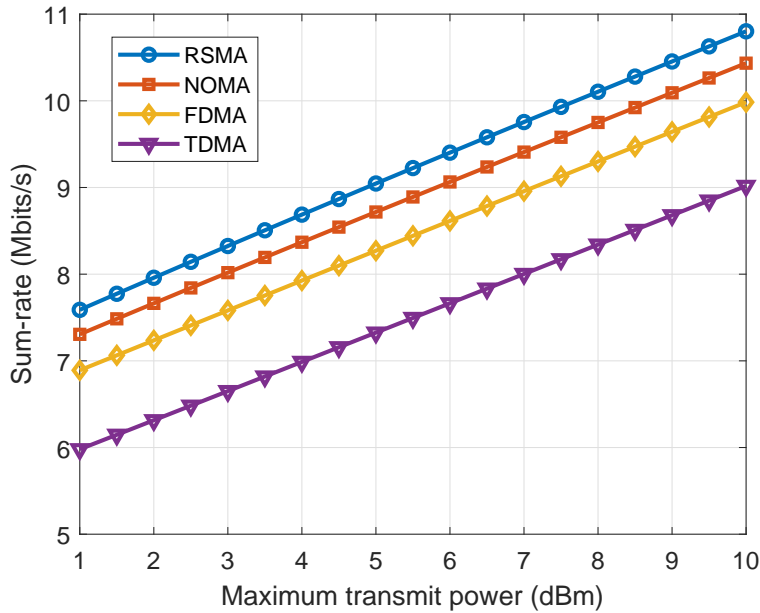}
\vspace{-0.5em}
\caption{Sum-rate  versus maximum transmit power of each user ($K=2$ users, $D_1=0.5$, and $D_2=0.5$).} \label{fig2}
\vspace{-0.5em}
\end{figure}

We compare the sum-rate performance of RSMA, NOMA, FDMA, and TDMA.
Fig. \ref{fig2} shows how the sum-rate changes as the maximum transmit power of each user varies for a network having two users.
We can see that the sum-rate of all multiple access schemes linearly increases with the logarithmic maximum transmission power of each user. This is because the sum-rate is a logarithmic function of the maximum power of the users.
It is found that RSMA achieves the best performance among all multiple access schemes.
From Fig.~\ref{fig2},
RSMA can increase up to  4.1\%, 10.2\% and 28.8\% sum-rate compared to NOMA, FDMA and TDMA, respectively.
This is because that RSMA can achieve the largest rate region and users with RSMA can achieve higher rate than other multiple access schemes.
Fig.~\ref{fig2} also shows that TDMA achieves the worst sum-rate performance, which corroborates the theoretical findings in Lemma 5.

\begin{figure}[ht]
\centering
\includegraphics[width=4.5in]{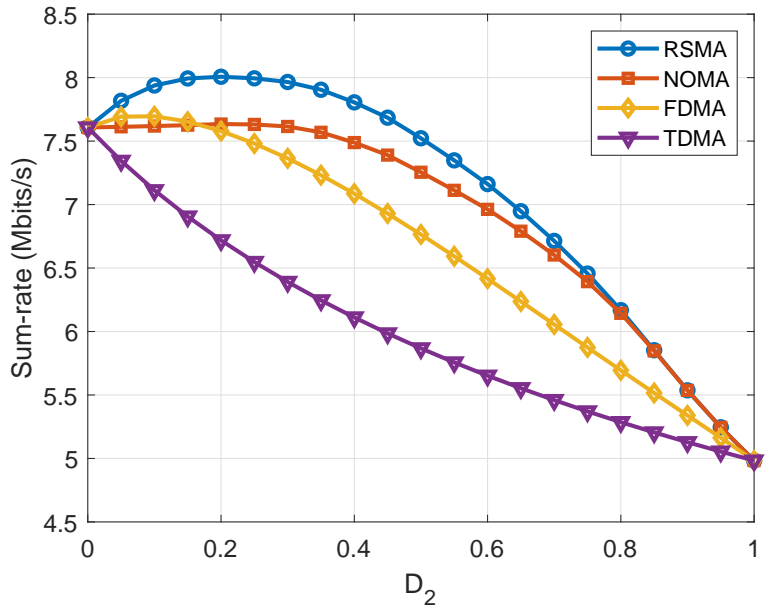}
\vspace{-0.5em}
\caption{Sum-rate versus proportional rate parameter $D_2$ ($K=2$ users and $D_1=1-D_2$).} \label{fig3}
\vspace{-0.5em}
\end{figure}

Fig. \ref{fig3} shows the sum-rate versus the proportional rate parameter $D_2$ of user 2.
From this figure, we can observe that RSMA achieves the best performance in terms of sum-rate, while TDMA has the worst sum-rate performance.
This is because RSMA exhibits a better spectrum efficiency compared to FDMA, and RSMA uses an optimized power allocation among two messages for each user to achieve the optimal rate region, while each user only transmits one message in NOMA.
In addition, both FDMA and NOMA achieve better sum-rate performance than TDMA due to the superiority of  spectrum efficiency.
Fig. \ref{fig3} shows that the sum-rate of NOMA is lower than that of FDMA for  small proportional rate parameter of user 2 ($D_2<0.2$) and the sum-rate of NOMA is higher than that of FDMA for proportional rate parameter of user 2 ($D_2>0.2$).
According to the proportional rate constraints (\ref{ad2minr0}a) $\frac{r_1}{r_2}= \frac{D_1}{D_2}$,  the optimal rate point can be obtained by finding the intersection between the rate region and the line  $r_2 =\frac{D_2}{D_1} r_1$.
For a small proportional rate parameter of user 2, the slope $\frac{D_2}{D_1}$ is low and user rate value of the intersection between the rate region (Fig. \ref{fig1}) of NOMA and  line  $r_2 =\frac{D_2}{D_1} r_1$ is smaller than that of the intersection between the rate region  of FDMA and line  $r_2 =\frac{D_2}{D_1} r_1$.
Consequently, the sum-rate of FDMA is higher than NOMA.
WBy using a similar, the sum-rate performance of NOMA is better than FDMA for large proportional rate parameter of user 2.

\begin{figure}[t]
\centering
\includegraphics[width=4.5in]{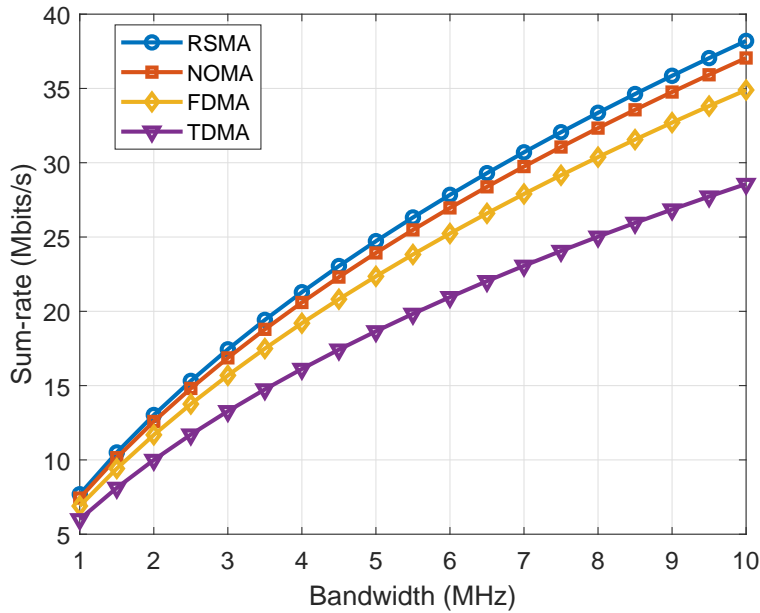}
\vspace{-0.5em}
\caption{Sum-rate versus bandwidth of the BS ($K=2$ users, $D_1=0.5$, and $D_2=0.5$).} \label{fig5}
\vspace{-0.5em}
\end{figure}

\begin{figure}[t]
\centering
\includegraphics[width=4.5in]{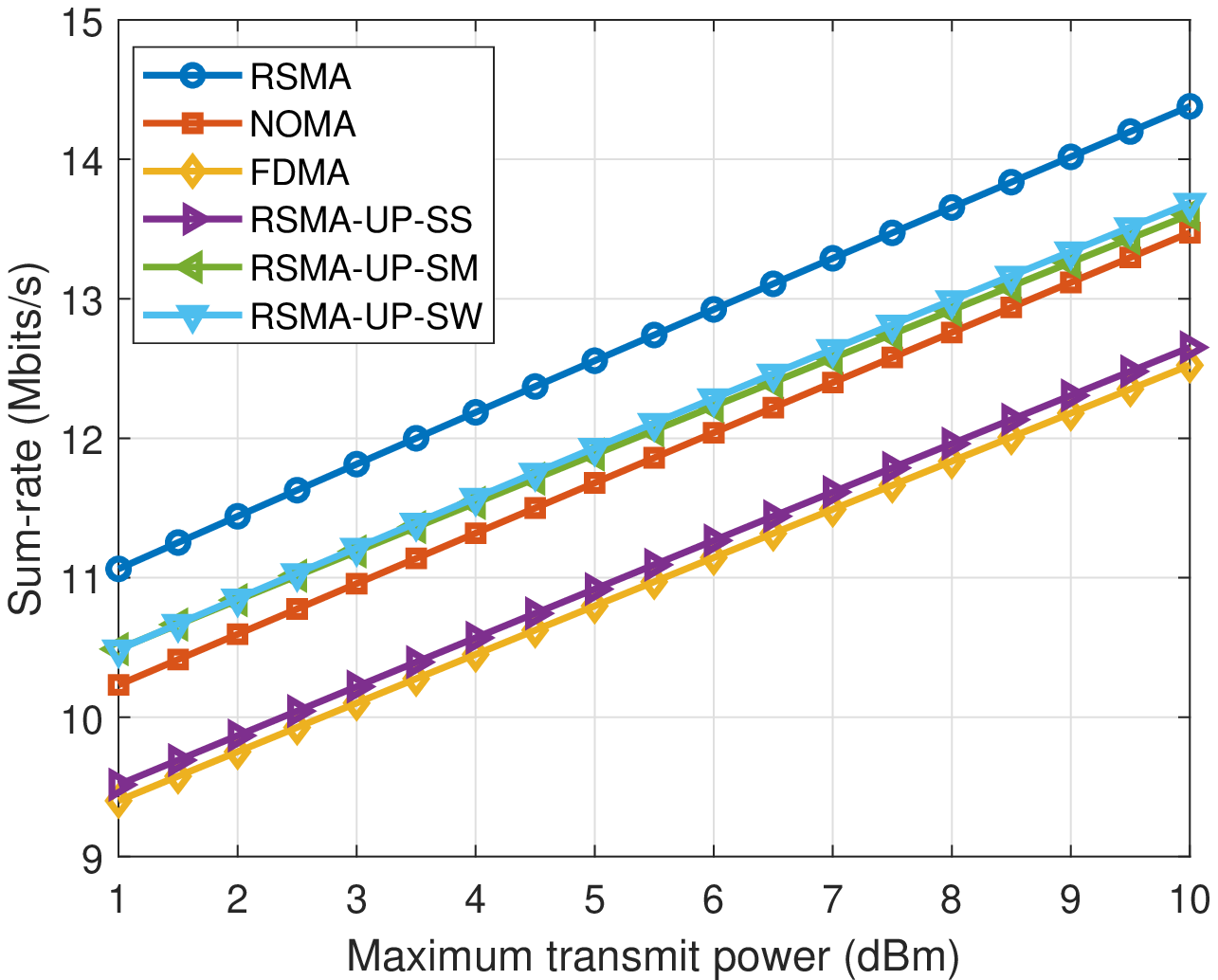}
\vspace{-0.5em}
\caption{Sum-rate versus maximum transmit power of each user under different user pairing metods ($K=10$ users, $D_1=\cdots=D_{10}=0.1$).} \label{fig7}
\vspace{-0.5em}
\end{figure}

Fig. \ref{fig5} shows the sum-rate versus the bandwidth of the BS.
From this figure, we can see that RSMA always achieves a better performance than NOMA, FDMA, and TDMA.
Fig. \ref{fig5} demonstrates that the sum-rate increases rapidly for a small bandwidth, however, this increase becomes slower for a larger bandwidth.
This is because a high bandwidth leads to high noise power, which consequently decreases the slope of increase of the sum-rate for all multiple access schemes.
Fig. \ref{fig5} also demonstrates that the sum-rate resulting from RSMA is greater than the one achieved by all other multiple access schemes, particularly when the bandwidth is large.


\begin{figure}[t]
\centering
\includegraphics[width=4.5in]{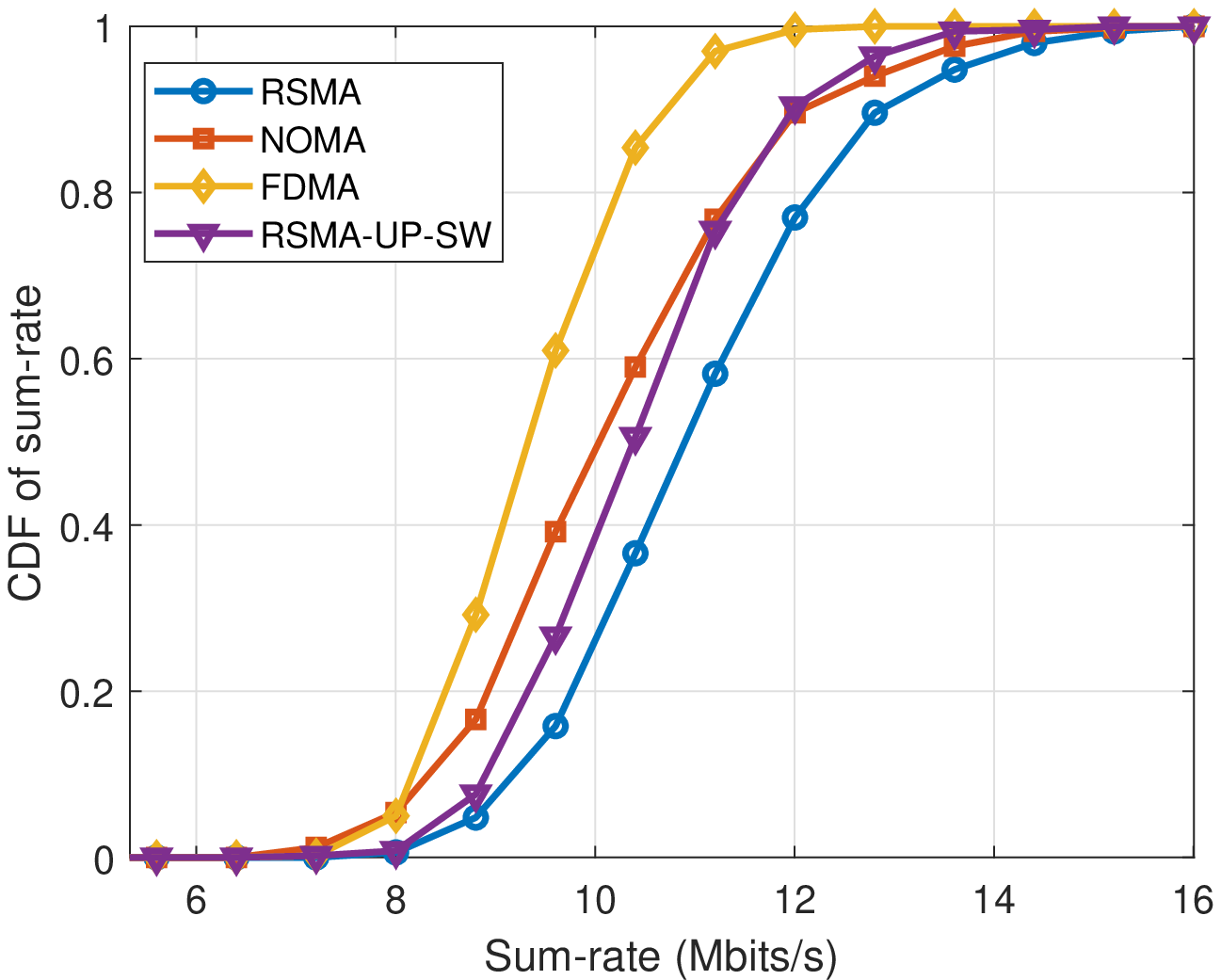}
\vspace{-0.5em}
\caption{CDF of sum-rate ($K=10$ users, $D_1=\cdots=D_{10}=0.1$).} \label{fig9}
\vspace{-0.5em}
\end{figure}

For low-complexity RSMA with user pairing (labeled as `RSMA-UP'),
we study the influence of user pairing by considering three different user-pairing methods \cite{7273963}.
For strong-weak (SW) pair selection, the user with the strongest channel condition is paired with the user with the weakest in one pair, and
the user with the second strongest is paired with one with the second weakest in one pair, and so on.
For strong-middle (SM) pair selection, the user with the strongest channel condition is paired with the user with the middle strongest user in one pair, and so on.
For strong-strong (SS) pair selection, the user with the strongest channel condition is paired with the one with the second strongest in one pair, and so on.

In Fig. \ref{fig7}, we  show how the sum-rate changes as the maximum transmit power of each user varies for a network having ten users.
From this figure, we observe that RSMA always achieves the best performance.
For RSMA-UP with different user-pairing methods, SW outperforms the other
two methods in terms of sum-rate for RSMA-UP.
To maximize the sum-rate, it tends to pair users with distinctive gains for sum-rate maximization.
Due to the superiority of SW, we choose SW for  pair selection of RSMA-UP in the following simulations.

Fig. \ref{fig9} presents the cumulative distribution function (CDF) of sum-rate resulting from RSMA, NOMA, FDMA, and RSMA-UP-SW for a network with $K=10$ users.
From Fig. \ref{fig9}, the CDFs for
RSMA, RSMA-UP-SW, and NOMA all improve significantly compared
FDMA, particularly when the sum-rate is high, which shows that RSMA, RSMA-UP-SW, and NOMA are suitable for high sum-rate transmission.
Moreover, we can observe that RSMA outperforms NOMA.
This is because RSMA can adjust the splitting power of two messages for each user so as to control the interference decoding thus optimizing the sum-rate of all users, while there is no power splitting for each user in NOMA.
Moreover, RSMA-UP-SW can achieve a similar performance to RSMA.
However, the complexity of RSMA-UP-SW  is much lower compared to RSMA according to Section III-B and Section IV,  which shows the effectiveness of  RSMA-UP-SW.

\begin{figure}[htb]
\centering
\includegraphics[width=4.5in]{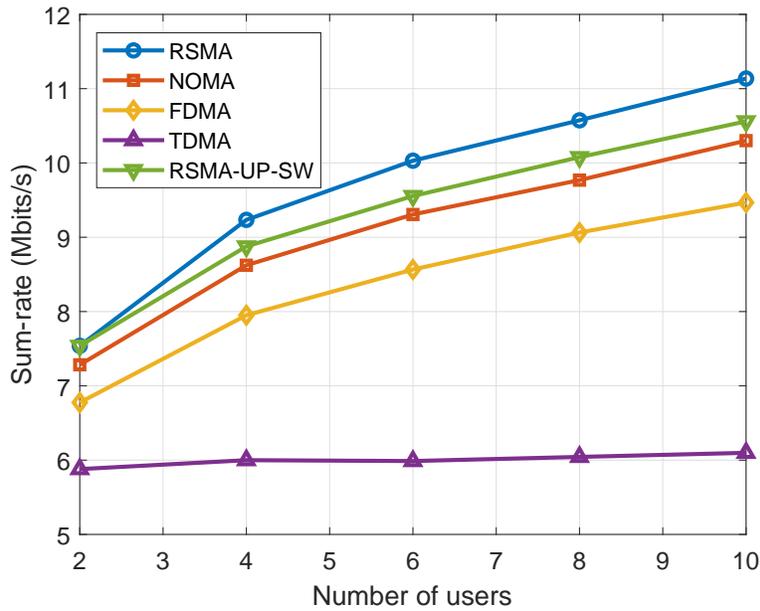}
\vspace{-0.5em}
\caption{Sum-rate versus number of users ($D_1=\cdots=D_{K}=1/K$).} \label{fig6}
\vspace{-0.5em}
\end{figure}

In Fig.~\ref{fig6}, we plot the sum-rate versus the number of users is given.
Clearly, the proposed RSMA or RSMA-UP-SW will always achieve a better performance compared to NOMA, FDMA, and TDMA especially when the number of users is large.
In particular, RSMA can achieve sum-rate gains of up to 10.0\%, 22.2\%, and 83.7\%  compared to NOMA, FDMA, and TDMA, respectively, while  RSMA-UP-SW can improve the sum-rate by up to 4.1\%, 11.6\% and 66.8\% compared to NOMA, FDMA, and TDMA, respectively.
When the number
of users is large, the multiuser gain is more pronounced for the proposed RSMA compared to conventional NOMA, FDMA, and TDMA.
This is due to the fact hat RSMA can effectively determine the power splitting of each user to achieve the theoretically maximal rate region, while there is no power splitting  in  NOMA and the allocated bandwidth/time of each user is low for FDMA/TDMA when the number of users is large.
However, RSMA achieves a better performance compared to NOMA, FDMA, and TDMA at the cost of additional computational complexity according to Section III-B.
Fig.~\ref{fig6} also shows that RSMA-UP-SW achieves  a better sum-rate performance compared to NOMA, FDMA, and TDMA but with low complexity according to Section IV, which shows that RSMA-UP-SW is promising solution that strikes a desirable tradeoff between performance gain and computational complexity.

\section{Conclusion}
In this paper, we have investigated the decoder order and power optimization in an uplink RSMA system. We have formulated the problem as a sum-rate maximization problem. To solve this problem, we have transformed it into an equivalent problem with only rate splitting variables, which has closed-form optimal solution.
Given the optimal rate requirement of each user, the optimal transmit power of each user is obtained under given the decoding order and the optimal decoding order is found by an exhaustive search method.
To reduce the computational complexity, we have proposed a low-complexity RSMA with user pairing.
Simulation results show that RSMA achieves higher sum-rate than NOMA, FDMA, and TDMA.

\appendices
\section{Proof of Lemma \ref{ad2minr1_2le2}}
\setcounter{equation}{0}
\renewcommand{\theequation}{\thesection.\arabic{equation}}
Assume that for the optimal solution $(\tau^*, \boldsymbol r^*)$ of problem (\ref{ad2minr1_2}), we have
\begin{equation}
 \sum_{k\in\mathcal K'}r_{k}^* < B \log_2 \left( 1+ \frac{\sum_{k\in\mathcal K'}h_kP_{k}}{\sigma^2B}\right), \quad \forall \mathcal K'\subseteq \mathcal K \setminus \emptyset.
\end{equation}

In this case, we can construct a new rate solution $\boldsymbol r'=[r_1',\cdots,r_K']$ with $r_k'=\epsilon r_k^* $ and
\begin{equation}\label{AppLema3eq1}
\epsilon\!=\!\min_{\mathcal K'\subseteq \mathcal K\setminus \emptyset}\!\frac  {B \log_2 \left( 1+ \frac{\sum_{k\in\mathcal K'}h_kP_{k}}{\sigma^2B}\!\right)}{\sum_{k\in\mathcal K'}r_{k}^*}   >1.
\end{equation}
According to (\ref{AppLema3eq1}), we
can show that
\begin{equation}\label{AppLema3eq2_1}
\sum_{k\in\mathcal K'}r_{k}' \leq B \log_2 \left( 1+ \frac{\sum_{k\in\mathcal K'}h_kP_{k}}{\sigma^2B}\right),\:\forall \mathcal K'\subseteq \mathcal K \setminus \emptyset,
\end{equation}
which ensures that $\boldsymbol r'$ satisfies constraints (\ref{ad2minr1_2}b).

Based on (\ref{ad2minr1_2}a), we have
$
\tau^*=  \frac{r_k^*}{D_k}, \forall k \in \mathcal K
$.
We set $\tau'$ as
\begin{equation}\label{AppLema3eq2}
\tau'= \frac{r_k'}{D_k}=\frac{\epsilon r_k^* }{D_k}>\tau^*.
\end{equation}

According to \eqref{AppLema3eq2_1} and \eqref{AppLema3eq2}, we can see that new solution $(\tau',\boldsymbol r')$ is feasible and the objective value (\ref{ad2minr1_2}) of new solution is better than that of solution $(\tau^*,\boldsymbol r^*)$, which contradicts the fact that  $(\tau^*,\boldsymbol r^*)$ is the optimal solution.
Lemma 2 is proved.

%

\section{Proof of Lemma \ref{ad2sedle1}}
\setcounter{equation}{0}
\renewcommand{\theequation}{\thesection.\arabic{equation}}

Given decoding orders $s_{21}$, $s_{11}$ and $s_{22}$, we have
\begin{equation}\label{Appad2sedeq1}
r_1=B\log_2\left(
1+\frac{h_1 p_{11}}{h_2 p_{22}+\sigma^2B}
\right)
\end{equation}
and
\begin{equation}\label{Appad2sedeq2}
r_{2}\!=\!B\log_2\!\left(\!
1\!+\!\frac{h_2 p_{21}}{h_1p_{11}\!+\!h_2p_{22}\!+\!\sigma^2B}
\!\right)\! +\!
B\log_2\!\left(\!
1\!+\!\frac{h_2 p_{22}}{  \sigma^2B}
\!\right).
\end{equation}


For case (1), since user 1 reaches its maximum rate point, we have $p_{11}=P_1$ and $p_{22}=0$. Transmission power $p_{21}$ can be calculated according to \eqref{Appad2sedeq2}.

For case (2), since user 2 reaches its maximum rate point, we have $p_{21}=0$ and $p_{22}=P_2$. Transmission power $p_{11}$ can be calculated according to \eqref{Appad2sedeq1}.

For case (3), since $r_1+r_2=R_{\max}$, the sum-rate of users 1 and 2 only reaches its maximum point when both users $u_1$ and $u_2$ transmit maximum power,  i.e., $p_{11}=P_1$ and $p_{21}+p_{22}=P_2$.
According to \eqref{Appad2sedeq1}, we can obtain power $p_{22}$.
With $p_{22}$, we can calculate $p_{21}=P_2-p_{22}$.

\section{Proof of Theorem \ref{ad2minn1th1}}
\setcounter{equation}{0}
\renewcommand{\theequation}{\thesection.\arabic{equation}}
Denote
\begin{equation}\label{AppThe2eq1}
z_k=\sum_{j=k}^{K}h_jq_j,\quad \forall k\in\mathcal K.
\end{equation}
According to (\ref{ad2minn1}a), we can obtain:
\begin{equation}\label{AppThe2eq2}
 B\log_2\left(
\frac{   z_{k}+\sigma^2B}
{   z_{k+1} +\sigma^2B}
\right)={D_k}{\tau}.
\end{equation}
Based on (\ref{AppThe2eq2}), we have:
\begin{equation}\label{AppThe2eq3}
z_{k}={2^{\frac{D_k\tau}{B}}} z_{k+1}  +\left({2^{\frac{D_k\tau}{B}}} -1\right)\sigma^2B.
\end{equation}

Using the recursive formulation (\ref{AppThe2eq3}) and $z_{K+1}=\sum_{k=K+1}^Kq_k=0$, we can calculate:
\begin{equation}\label{AppThe2eq5}
z_{k}= \sum_{j=k}^K 2^{\frac{\sum_{l=k}^{j-1} D_l\tau}{B}}\left({2^{\frac{D_j\tau}{B}}} -1\right)\sigma^2B,
\end{equation}
where we set $\sum_{l=k}^{k-1} D_l=0$.
Based on \eqref{AppThe2eq1}, we have:
\begin{equation}\label{AppThe2eq6}
q_k=\frac{z_{k}-z_{k+1}}{h_k}, \quad\forall k\in \mathcal K.
\end{equation}
Combining \eqref{AppThe2eq5} and \eqref{AppThe2eq6} yields:
\begin{equation}\label{AppThe2eq7}
q_k=\frac 1 {h_k}{\left(2^{\frac{D_k\tau}{B}}-1\right)} \sum_{j=k+1}^K 2^{\frac{\sum_{l=k+1}^{j-1} D_l\tau}{B}}\left({2^{\frac{D_j\tau}{B}}} -1\right)\sigma^2B +\frac 1 {h_k}\left({2^{\frac{D_k\tau}{B}}} -1\right)\sigma^2B, \quad\forall k\in \mathcal K,
\end{equation}
which monotonically increases with $\tau$.
Considering the maximum uplink transmission power constraints (\ref{ad2minn1}b),
we can obtain that
\begin{equation}\label{AppThe2eq7_2}
\tau \leq \tau_k, \quad\forall k\in \mathcal K,
\end{equation}
where $\tau_k$ is the solution to \eqref{ad2minn1th1eq3}.
To maximize sum-rate $\tau$, the optimal $\tau^*$ is given by \eqref{ad2minn1th1eq1}.
Consequently, the optimal transmission power is provided by \eqref{ad2minn1th1eq2}.

\section{Proof of Theorem \ref{ad2minf1th1}}
\setcounter{equation}{0}
\renewcommand{\theequation}{\thesection.\arabic{equation}}

Solving (\ref{ad2minf1}a) yields
\begin{equation}\label{Appad2minf1th1eq2}
b_k=-\frac{(\ln2)D_kh_kP_k\tau }
{Bh_kP_k W\left(-\frac{(\ln2)D_k\sigma^2\tau}{h_kP_k} \text e^{-\frac{(\ln2)D_k\sigma^2\tau}{h_kP_k}}
\right)+ {(\ln2)D_k\sigma^2B\tau}
}.
\end{equation}
 Substituting \eqref{Appad2minf1th1eq2} to constraint (\ref{ad2minf1}b), the optimal $\tau^*$ is calculated as \eqref{ad2minf1th1eq1}, and the optimal bandwidth allocation is accordingly given by \eqref{ad2minf1th1eq2}.

\section{Proof of Theorem \ref{ad2mint1th1}}
\setcounter{equation}{0}
\renewcommand{\theequation}{\thesection.\arabic{equation}}

Solving (\ref{ad2mint1}a) yields
\begin{equation}\label{Appad2mint1th1eq2}
a_k= \frac{ D_k \tau }
{ B\log_2 \left( 1+ \frac{h_kP_k}{\sigma^2B}
\right)
}.
\end{equation}
Combining \eqref{Appad2mint1th1eq2} and constraint (\ref{ad2mint1}b), the optimal sum-rate is given by \eqref{ad2mint1th1eq1}.
Then, the optimal time sharing is  \eqref{ad2mint1th1eq2}.

\section{Proof of Lemma \ref{ad2sedle2}}
\setcounter{equation}{0}
\renewcommand{\theequation}{\thesection.\arabic{equation}}

For two-user NOMA, according to \eqref{sys1eq10}, we can obtain:
\begin{equation}\label{Appad2sedle2eq1}
r_1=B\log_2\left(
1+\frac{h_1q_1}{h_2q_2+\sigma^2B}
\right),
r_2=B\log_2\left(
1+\frac{h_2q_2}{\sigma^2B}
\right),0\leq q_1\leq P_1,0\leq q_2\leq P_2.
\end{equation}
Based on \eqref{Appad2sedle2eq1}, we have:
\begin{equation}\label{Appad2sedle2eq2}
r_2=B\log_2\left(
1+\frac{h_2q_2}{\sigma^2B}
\right)\leq B\log_2\left(
1+\frac{h_2P_2}{\sigma^2B}
\right)= R_2,
\end{equation}
and
\begin{equation}\label{Appad2sedle2eq3}
h_2q_2+\sigma^2B=\frac{h_1 q_1}{2^{\frac{r_1}{B}}-1}.
\end{equation}
Substituting \eqref{Appad2sedle2eq3} into \eqref{Appad2sedle2eq2} yields:
\begin{align}\label{Appad2sedle2eq5}
r_2&=B\log_2\left(
1+\frac{h_2q_2}{\sigma^2B}
\right)
=B\log_2\left(
\frac{h_1 q_1}{\left(2^{\frac{r_1}{B}}-1\right)\sigma^2B}
\right)
\nonumber\\
&=B\log_2\left(
\frac{h_1 q_1}{\sigma^2B}
\right)-B\log_2\left( { 2^{\frac{r_1}{B}}-1 }
\right)
\leq B\log_2\left(
\frac{h_1 P_1}{\sigma^2B}
\right)-B\log_2\left( { 2^{\frac{r_1}{B}}-1 }
\right),
\end{align}
where the inequality follows from the fact that $0\leq q_1\leq P_1$.
Combining \eqref{Appad2sedle2eq1} to \eqref{Appad2sedle2eq5}, two-user NOMA rate region is given in \eqref{ad2sedle2eq1}.

For two-user FDMA, according to \eqref{sys1eq11}, we have:
\begin{equation}\label{Appad2sedle2eq6}
r_1=Bb_1 \log_2 \left( 1+ \frac{h_1P_1}{\sigma^2Bb_1}
\right),  r_2=Bb_1 \log_2 \left( 1+ \frac{h_2P_2}{\sigma^2Bb_2}
\right),
b_1+b_2=1, b_1,b_2 \geq 0.
\end{equation}
Based on \eqref{Appad2sedle2eq6},
we have
\begin{equation}\label{Appad2sedle2eq7}
b_k=f_k(r_k), \quad k=1,2,
\end{equation}
where $f_k(r_k)$ is defined in \eqref{ad2sedle2eq5}.
Combining \eqref{Appad2sedle2eq6} and \eqref{Appad2sedle2eq7}, two-user FDMA rate region is given in \eqref{ad2sedle2eq2}.

For two-user TDMA, according to \eqref{sys1eq12}, we can calculate:
\begin{equation}\label{Appad2sedle2eq8}
r_1=a_1 R_1,  r_2=a_2  R_2,
a_1+a_2=1, a_1,a_2 \geq 0.
\end{equation}
According to \eqref{Appad2sedle2eq8}, two-user TDMA rate region is given in \eqref{ad2sedle2eq3}.

\vspace{-0.5em}
\bibliographystyle{IEEEtran}
\bibliography{IEEEabrv,MMM}

\end{document}